# Machine learning applications for weather and climate need greater focus on extremes

Peter AG Watson[1,2]

1. School of Geographical Sciences, University of Bristol, Bristol, UK.
2. Cabot Institute for the Environment, University of Bristol, Bristol, UK.

**Introduction**

Multiple studies have now demonstrated that machine learning (ML) can give improved skill for predicting or simulating fairly typical weather events, for tasks such as short-term and seasonal weather forecasting (Ham et al., 2019; e.g. Ravuri et al., 2021; Weyn et al., 2021; Pathak et al., 2022), downscaling simulations to higher resolution (e.g. Stengel et al., 2020; Harris et al., 2022) and emulating and speeding up expensive model parameterisations (e.g. Rasp et al., 2018; Gettelman et al., 2021). These used ML methods with very high numbers of parameters, such as neural networks, which are the focus of the discussion here. Not much attention has been given to the performance of these methods for extreme event severities of relevance for many critical weather and climate prediction applications. This leaves a lot of uncertainty about the usefulness of these methods, particularly for general purpose prediction systems that must perform reliably in extreme situations. ML models may be expected to struggle to predict extremes due to there usually being few samples of such events. However, as will be discussed below, there are some studies that do indicate that ML models can have reasonable skill for extreme weather, and that it is not hopeless to use them in situations requiring extrapolation. This makes it an area worth researching more.

Some clarity is needed about the use of the term "extreme". One useful metric to represent the degree to which an event is extreme is the return period, the average time between events with a magnitude at least as large as for the event in question. A large number of studies use the term "extreme" to describe events around the 90-99$^{th}$ percentile of daily data, which correspond to only a 10-100 day return period. It is indeed useful to assess the performance of ML models around such thresholds. However, these are far from event severities that are relevant to many applications of weather and climate models, and studies typically do not demonstrate how their methods would perform in these cases.

At the high end of the scale, events with return periods in the thousands of years are sometimes studied in extreme event attribution (e.g. Risser & Wehner, 2017; Van Oldenborgh et al., 2017) and in the hundreds of years for designing infrastructure for flood and drought resilience (e.g. Environment Agency, 2014, 2020). In weather forecasting, the Met Office's most severe "red" weather warning was issued once every few years per event type in the system's first decade (Suri & Davies, 2021). The return period at individual locations that were most affected by these events will have been substantially higher.

Forecast reliability will also need to be assured for even more extreme events. In keeping with these examples, in the rest of this article "extreme" is used to refer to events with return periods of more than a few years.

It seems likely that for ML-based systems to be considered for use in operational weather and climate prediction systems, good performance in extreme situations needs to be shown. This should include events going beyond what is used for training systems, since it cannot be known in advance what range of input data the system will see. Operational systems need to predict events that are more severe than any in the historical record at times. It can be asked is there much value in continuing development of ML-based systems for weather and climate prediction without demonstrating at least satisfactory performance for extremes?

If an approach is taken to try to first design systems to perform well for typical weather and then improve extreme event capabilities later, this could waste a lot of time if useful methods for the former are not the same as for the latter. This is an especially large concern for ML methods with large numbers of parameters (e.g. large neural networks) that require a lot of samples for training. Particular methods may also have their own vulnerabilities. For example, generative adversarial networks are prone to "mode collapse", where predictions seriously undersample parts of the data distribution, potentially very adversely affecting performance for extremes. Random forests cannot predict values beyond those seen in training data, so they may not be a good choice for applications where skilful prediction for beyond-sample events is important. Therefore evaluating how well such systems actually perform in extreme situations is very important for helping researchers choose the best methods to develop for their applications.

The challenge in making predictions in extreme situations comes not just from these events being rare, but also from how far they can exceed historical records. The 2021 heatwave in the Northwest USA and western Canada beat previous temperature records by 5°C in Portland, standing far above previous values, with an estimated return period in the present climate of ~1000 years (Philip et al., 2021). Climate model simulations include events where weekly-average temperature exceeds previous records by over five standard deviations (Fischer et al., 2021). Rainfall extremes can exceed prior historical values by even greater margins. In 2018 and 2019 in Kerala, India, there were 14-day rainfall totals that exceeded 30 standard deviations, associated with strong convection (Mukhopadhyay et al., 2021). Convective rainfall in the USA has led to river discharges reaching over 20 times the 10-year return level on a large number of occasions, with the most extreme recorded discharge due to rainfall being 200 times that level (Smith et al., 2018). It therefore wouldn't be over the top to evaluate robustness of ML-based systems to this degree of extremity for cases where convection is important, and otherwise to perhaps ~5 standard deviation perturbations above the highest values in observed or simulated training data.

**Previous studies evaluating ML on extreme events**

There are six studies that I have been able to find in the literature that indicate that ML-based systems can have reasonable skill in extreme situations with return periods of more than a few years. These are summarised in table 1.

**Table 1:** Summary of six studies that found that ML-based systems can perform reasonably at predicting extreme events that have return periods of more than a few years.

| Study | ML Method | Training dataset size | Maximum return period evaluated | Notes |
|---|---|---|---|---|
| Adewoyin et al. (2021) | Convolutional recurrent neural network | 10 years | 6 years | • Downscaled daily-mean precipitation at 16 UK locations. |
| Boulaguiem et al. (2022) | Generative adversarial network (GAN) | 50 years | ~2000 years | • Produced samples of maps of annual summer maximum temperature and winter maximum precipitation over Europe.<br>• The density in the tails of the predicted distribution appeared reasonable, though errors were not precisely quantified.<br>• The structure of their GAN was adapted to work better for extremes. |
| Frame et al. (2022) | Long short-term memory neural network | Up to 34 years per river catchment | >100 years | • Predicted river flows in the USA.<br>• In one test they removed events in the training dataset with return periods greater than 5 years and found that prediction scores were still good for events with return periods exceeding 100 years (estimated using a fitted distribution). |
| Grönquist et al. (2021) | Convolutional neural network | 15 years | Unquantified, but record-breaking | • Postprocessed global weather forecasts at 48 hour lead time.<br>• Improved forecast skill scores on extreme events including Hurricane Winston (the most intense |

| | | | | southern hemisphere hurricane on record) and an unprecedented cold wave in southeast Asia. |
|---|---|---|---|---|
| Lopez-Gomez et al. (2022) | Convolutional neural network | 24 years | ~1000 years | • Global weather forecasts of daily temperature, up to lead times of 4 weeks.<br>• Produced sensible forecasts for record-breaking events: the 2017 European heatwave and the 2021 Northwest USA heatwave.<br>• They used a modified loss function that put greater weight on extreme events. |
| Nevo et al. (2021) | Bespoke combination of ML models | 5 years | 5 years | • Predicted flooding in India and Bangladesh.<br>• To evaluate the performance of their flood inundation model on extreme events, they targeted the most severe event in each river basin in a 5-year dataset. Events with inundation level within 30cm of the target event were removed, and the model was trained on the remaining data, making this a test of performance on unprecedented events as far as the model knew.<br>• The median performance on these events was similar to that for typical events, though the skill for a few of the extreme events was poor. |

These results show that there are good prospects that ML-based systems could have good skill for extreme events with multi-year return periods and beyond, but there are not enough studies to know whether this is true in most cases. I have not found any studies that explicitly show failure for extremes. It is hard to draw general rules for success from this small sample of results, but it does suggest some guidance. Five out of six studies evaluated neural network-based models, indicating that neural networks can be successful for this task. The other study, Nevo et al. (2021), used a bespoke approach for their flood inundation modelling. No study tested alternative complex methods like random forests, so they provide no evidence about such methods. Five out of six studies used at least 10 years of training data. Three studies obtained reasonable evaluation results for extreme events with estimated return periods much longer than the training dataset, indicating that generalisation to more extreme events is possible (Boulaguiem et al., 2022; Frame et al., 2022; Lopez-Gomez et al., 2022). However, it still seems wise to plan to require large training datasets to develop models for such cases. Four out of six studies did not change their model architecture or training procedure to particularly target achieving good performance on extremes, again suggesting that existing methods are capable of generalising to extreme events, but modifications are sometimes needed.

**Research gaps**
More research into any aspect of this problem would be very valuable, though there are some questions that need answers with higher urgency. One highly important area is simulating weather events with multi-decadal return periods, which are very important for understanding many aspects of climate risk. Another is simulating situations that are multiple standard deviations beyond historical records. This has only been examined by Lopez-Gomez et al. (2022). Another key gap is testing how well stochastic generative models (e.g. generative adversarial networks), which have become popular for high-resolution downscaling and forecasting, perform in extreme situations. The challenge for these may be even greater than for deterministic models, and it has only been studied by Boulaguiem et al. (2022).

There is also a lack of studies that have examined ML models' extrapolation behaviour. Neural networks' extrapolation properties depend on their structure e.g. those using the common "ReLU" activation function would be expected to extrapolate linearly, though not necessarily with the same gradient as a line of best fit through the training data points (Xu et al., 2020; Ziyin et al., 2020). Hernanz et al. (2022) examined extrapolation behaviour of ML models that predicted surface air temperature and found that they performed poorly. Extrapolation errors may not strongly affect skill scores for events that do not lie very far outside the training data range. This makes it unclear if the models in the studies in table 1 contained this error. This kind of error would be more important for extremes far outside the training range (e.g. Fischer et al., 2021; Mukhopadhyay et al., 2021; Philip et al., 2021).

**Ways forward**
Firstly, studies could include diagnostics that indicate performance on extremes without requiring much extra work. For example:
- Scatter plots of predictions versus truth values, which immediately show whether an ML model predicts sensible values in the most extreme situations in the test data, and how prediction skill for extremes compares to more typical situations (as shown in e.g. Adewoyin et al., 2021, fig. 7).

- Quantile-quantile plots including percentiles corresponding to the highest allowed by the test data, which would greatly help to show whether the frequency of extreme events in predictions is reasonable.
- When predicting a spatial field in two or more dimensions (e.g. in downscaling), showing that predictions for samples of the most extreme cases in the test data are sensible.
- Statistics like root mean square error and correlation for the most extreme events only (e.g. the top 30 events). When these scores are calculated on a whole dataset, they are not very sensitive to errors in the distribution tails.
- Making clear in the conclusions what is the maximum return period of events that were evaluated in the test data.

To show how well ML-based systems perform in situations going beyond events seen in training, the most extreme events can be set aside in a second test dataset, as in Frame et al. (2022) and Nevo et al. (2021). This approach could be made even stronger by doing this before any model development is done, so the model structure and hyperparameters are chosen without being able to see the most extreme events beforehand.

It would also be highly valuable to understand how ML-based systems would perform in situations that are far out-of-sample, addressing the extrapolation question. For certain applications, increasing the magnitude of anomalies in input fields would be expected to result in increased magnitudes of anomalies in predicted values (e.g. in downscaling, parameterisation emulation, short-range forecasting). Then it would be very useful to show how the predictions scale as anomalies in input fields are magnified to correspond to events much more severe than any in the source data, up to multiple standard deviations beyond the sample events. It may improve confidence in the system if the predictions varied smoothly, if there is no reason to expect a sharp change.

Trustworthiness of predictions of extremes may also be informed by quantifying uncertainty associated with model structure and parameters (e.g. Abdar et al., 2021) and interpretability methods (e.g. McGovern et al., 2019; Ebert-Uphoff & Hilburn, 2020; Toms et al., 2020; Beucler et al., 2022). However, the reliability of interpretability methods has been questioned (e.g. Lipton, 2018; Rudin, 2019; Koch & Langosco, 2021). I am not aware of tests of these approaches on predictions of extreme events, and these would be very valuable.

If existing machine learning approaches turn out not perform well enough at predicting extreme events, this would signal that more effort should be put into designing systems that are robust. For example, physical principles could be incorporated (e.g. Beucler et al., 2021) or systems that are hybrids of conventional and ML-based models could be developed, which may be more reliable (e.g. Watson, 2019; Bonavita & Laloyaux, 2020; Brajard et al., 2021). For emulating expensive conventional models, rare event simulation (Ragone et al., 2017; Webber et al., 2019) could be useful for obtaining sufficiently many extreme events for training. Better diagnostics of performance on extreme events in studies applying ML would be very valuable for determining whether more attention should turn to approaches like these.

**Conclusions**

In order for ML to be applied broadly in weather and climate prediction and simulation systems, it needs to be shown that it can perform at least reasonably well for extreme events. ML models with high numbers of parameters, such as neural networks, may be expected to struggle in these cases as they typically need large samples of events to be trained to make skilful predictions. However, the six studies reviewed here that do evaluate ML model skill on extremes actually indicate that ML-based systems can still perform well on out-of-sample extreme events, even for those with return periods of hundreds or thousands of years. This sample of studies is not enough to draw general conclusions from, though, and there are important questions that have not been addressed by any study that I could find. The situation could be greatly improved if study authors added certain simple diagnostics, and also if studies were designed to show the performance for extremes, as described above. This would be highly valuable for the rest of the community who would learn what ML methods are best to use to predict and simulate extreme events successfully.

**Acknowledgements**

I thank Peter Dueben for comments on an earlier draft of this manuscript. I also thank the editor and two anonymous reviewers for helpful comments. This work was supported by a NERC Independent Research Fellowship (grant no. NE/S014713/1).